\documentclass[%
 reprint,
superscriptaddress,
 amsmath,amssymb,
 aps,
 prx,
]{revtex4-2}

\usepackage{amsthm}

\usepackage{graphicx}
\usepackage{dcolumn}
\usepackage{bm}
\usepackage{physics}
\usepackage{booktabs}
\usepackage{enumerate}

\usepackage{lipsum} 

\usepackage[pdftex,pagebackref=false]{hyperref} 
\hypersetup{
    colorlinks=true,        
    linkcolor=cyan,         
    citecolor=cyan,        
    filecolor=magenta,      
    urlcolor=cyan           
}

\usepackage{xcolor}
\usepackage{cleveref}
\usepackage{placeins}
\usepackage{orcidlink}
\renewcommand\vec{\boldsymbol}

\newcommand{\aalto}{Department of Applied Physics, Aalto University, P.O. Box 15100, FI-00076 AALTO, Finland}
\newcommand{\lmu}{Fakultät für Physik, Ludwig-Maximilians-Universität, München, Germany}
\newcommand{\lmuu}{Munich Center for Quantum Science and Technology (MCQST), München, Germany.}
\newcommand{\delft}{Kavli Institute of Nanoscience, Delft University of Technology, 2628 CJ Delft, The Netherlands}
\newcommand{\leiden}{Leiden Institute of Physics, Leiden University, Niels Bohrweg 2, 2333 CA Leiden, The Netherlands}
\newcommand{\jyvaskyla}{Department of Physics and Nanoscience Center, University of Jyväskylä, P.O. Box 35 (YFL), FI-40014 Jyväskylä, Finland}
\newcommand{\maryland}{Maryland Quantum Materials Center, Department of Physics, University of Maryland, College Park, Maryland 20742-4111, USA}

\usepackage{multirow}

\newcommand\commastsym{\raisebox{-.5ex}{\shortstack{%
  \(\circ\)\\[-.5ex]%
  \(,\)}}%
}
\newcommand\commast{\protect\commastsym}

\begin{document}

\title{In situ replaceable matching circuits for quantitative microwave readout of the scanning tunneling junction}
\author{Amber Mozes \orcidlink{0009-0001-5694-4321}}
\affiliation{\leiden}
\affiliation{\aalto}
\author{Sanghun Lee \orcidlink{0009-0001-4930-1037}}
\affiliation{\lmu}
\affiliation{\lmuu}
\author{Suzanne van Noordt}
\affiliation{\leiden}
\author{Shaojie Yang}
\affiliation{\lmu}
\affiliation{\lmuu}
\author{Tjerk Benschop}
\affiliation{\leiden}
\author{Koen M. Bastiaans}
\affiliation{\delft}
\author{Tim Kokkeler \orcidlink{0000-0001-8681-3376}}
\affiliation{\jyvaskyla}
\author{Tero T. Heikkilä \orcidlink{0000-0002-7732-691X}}
\affiliation{\jyvaskyla}
\author{Steven M. Anlage}
\affiliation{\maryland}
\author{Milan P. Allan \orcidlink{0000-0002-5437-1945}}
\affiliation{\lmu}
\affiliation{\lmuu}
\affiliation{\leiden}

\begin{abstract}
Scanning tunneling microscopy (STM) reaches atomic resolution because the tunneling current (the resistive, dissipative channel of the tip–sample junction) is exponential in tip–sample distance. Performing STM at microwave frequencies would give access to the local complex impedance, but it brings a challenge: a microwave signal couples most strongly to the reactive, capacitive part of the tip–sample load, whose length scale is set by the much larger tip radius, while the dissipative part, a high impedance poorly matched to a 50 $\Omega$ line, is hard to reach. Here we combine STM with microwave reflectometry using a bias tee and matching circuit mounted directly behind the tip on an in situ replaceable “tip-chip”. The chip keeps the stray capacitance at 0.5 pF and lets us compare several matching circuits on the same microscope. A single equivalent-circuit model, with only two parameters that vary between chips, describes all circuits. The reactive and dissipative parts of the load appear largely in different features of the matched resonance: the capacitance predominantly shows up in the phase and yields an effective tip radius of 100 nm, whereas the junction resistance changes its depth in the magnitude of the S parameters as predicted from the DC setpoint. With the DC bias line disconnected, the microwave response alone images the Au(111) herringbone reconstruction, resolving height changes of about 20 pm. Across a monatomic step, the contrast is more than three orders of magnitude larger than the capacitive channel can produce and depends on the junction setpoint as the model predicts. The in situ replaceable tip-chip thus enables flexible, impedance-matched microwave STM, and we propose it as a route to measuring the local impedance of inhomogeneous conductors and superconductors.
\end{abstract}
\maketitle

\section{Introduction}
Measuring the local complex impedance yields information about a sample that its DC conductance does not provide. This matters for unconventional and inhomogeneous superconductors, where the reactive and dissipative parts of the response, the kinetic inductance and the quasiparticle loss, can differ between regions only nanometers apart \cite{
Kamplapure2013,Cho2019,Tromp2023}. Local maps of both could help decide whether macroscopic superconductivity in such materials is suppressed by weak pairing or by a loss of phase stiffness \cite{Emery1995,Crane2007}. In principle, one could measure the complex impedance locally by sending a microwave field to the sample through a sharp tip and recording the reflected wave. That reflection is set by the tip–sample load, which depends on the tip–sample distance, the tip radius, and the electromagnetic properties of the sample \cite{Xiang2002,Barber2022,Anlage2007}. However, in the near-field regime the load is dominated by the reactive coupling between tip and sample. For a tip–sample distance $d$ smaller than the tip radius $r_\mathrm{tip}$, the local response is measured over a length scale set by the tip diameter \cite{Mueller1994,Cory1998,Kurokawa1998,Imtiaz2003,Imtiaz2005,Lee2010,Reznik2014,Voogd2017}. It is usually described by the tip–sample capacitance that scales roughly $C_\mathrm{ts}\propto r_\mathrm{tip} \ln{(8r_\mathrm{tip}/d)}$ and can be modelled with a finite-element treatment of the tip geometry \cite{Nguyen2015,Jelic2017,Chu2020,Lee2020,Visser2016,Berweger2022,Lai2009,Gao2005,Barber2022}, which varies from tip to tip \cite{Imtiaz2006}. The resolution is then tied to the tip radius, typically tens of nanometers, and the dissipative part of the response is only a small correction on a large capacitance.

The tunnel junction adds a resistance $R_\mathrm{J}$, in parallel with $C_\mathrm{ts}$ \cite{Reznik2014,Kemiktarak2007}, which is large, of order 0.1 to 100 $\mathrm{M}\Omega$, and strongly mismatched from the 50 $\Omega$ of the microwave line. Coupling to it means matching the source to a load very different from its characteristic impedance. Different works have shown that it is possible to measure the tunneling response at microwave frequencies \cite{Imtiaz2006,Kemiktarak2007,Marbey2026} and that atomic resolution is achieved when measuring in the tunneling regime \cite{Lee2010}. However, the contributions to the tip–sample impedance have not been separated quantitatively in this regime. Without such a separation, a change in the junction impedance cannot be distinguished from a change in the sample impedance. In addition, microwave impedance microscopes typically use a fixed matching circuit, which limits them to the impedance range it was designed for \cite{Barber2022}.

Here we address both issues with an in situ replaceable tip-chip that lets us separate the reactive and dissipative channels. We place the bias tee and the matching circuit on a small chip directly behind the STM tip. This keeps the stray capacitance low, and because the chip can be swapped in situ, we can test many matching circuits on the same microscope. A single equivalent-circuit model, with the surface-mount components based on their datasheets and adapted to cryogenic temperatures, describes every circuit and lets us follow the tip–sample load as a function of distance and split it into a reactive and a dissipative part. Through the dissipative part we read out the tunnel junction and image the Au(111) surface with about 20 pm height sensitivity from the microwave response alone, with the DC bias disconnected, in agreement with the corresponding topography. As a caveat, we do not yet extract a separate sample impedance $Z_\mathrm{s}$.

\section{Setup and equivalent circuit model}

\begin{figure}
    \centering
    \includegraphics[width=\linewidth]{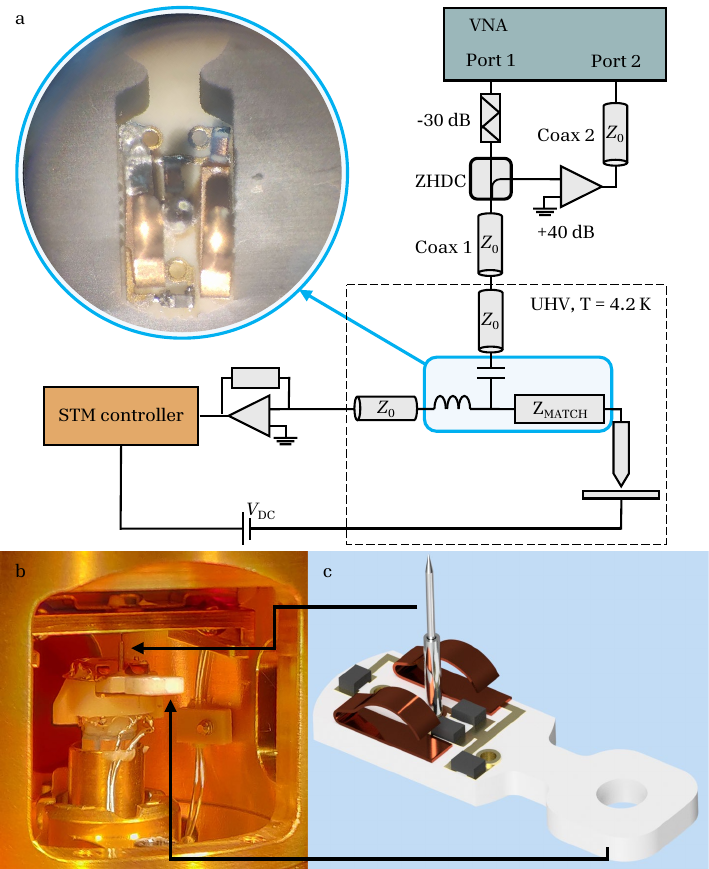}
    \caption{a) Measurement setup combining microwave read-out with a vector network analyzer (VNA, top in teal) and low frequency STM/STS (bottom left in orange). Inside the cryostat (UHV, T = 4.2 K), two cables are connected to the bias tee and impedance matching circuit that are implemented right before the STM tip (in blue rectangle). b) Picture of the STM head showing the two cables and impedance matching chip with STM tip. c) In situ replaceable chip circuits containing the circuit shown in blue rectangle (and picture inset) in a. Top arrow points at the STM tip, bottom arrow at the replaceable tip-chip.}
    \label{fig:fig1}
\end{figure}

We combine standard STM and spectroscopy with microwave reflectometry (\hyperref[fig:fig1]{Fig. \ref*{fig:fig1}}). The STM applies a DC bias voltage and measures the tunneling current, while a vector network analyzer (VNA) generates the microwave signal and measures the complex reflection coefficient $S_\mathrm{11}$. Both signals reach the STM tip simultaneously and are separated by a bias tee on our tip-chip. We compare different matching networks by exchanging the chip in situ while keeping the same tip, sample, and wiring.

Our tip-chips are custom design made from Rogers material RO4003 with copper conducting paths ordered from Eurocircuits. We make use of the coaxial wiring in the STM head (\hyperref[fig:fig1]{Fig. \ref*{fig:fig1}b}) which includes two cables, connected to three pins and the tip shield. Two pins are used for the DC and AC lines, whereas the third pin and the tip shield are connected to the ground of each line respectively. The springs clamp the chip inside the STM head and make electrical contact with the tip shield; to connect this ground to the DC cable, they are also connected to the third pin with silver epoxy.

\begin{figure}
    \centering
    \includegraphics[width=\linewidth]{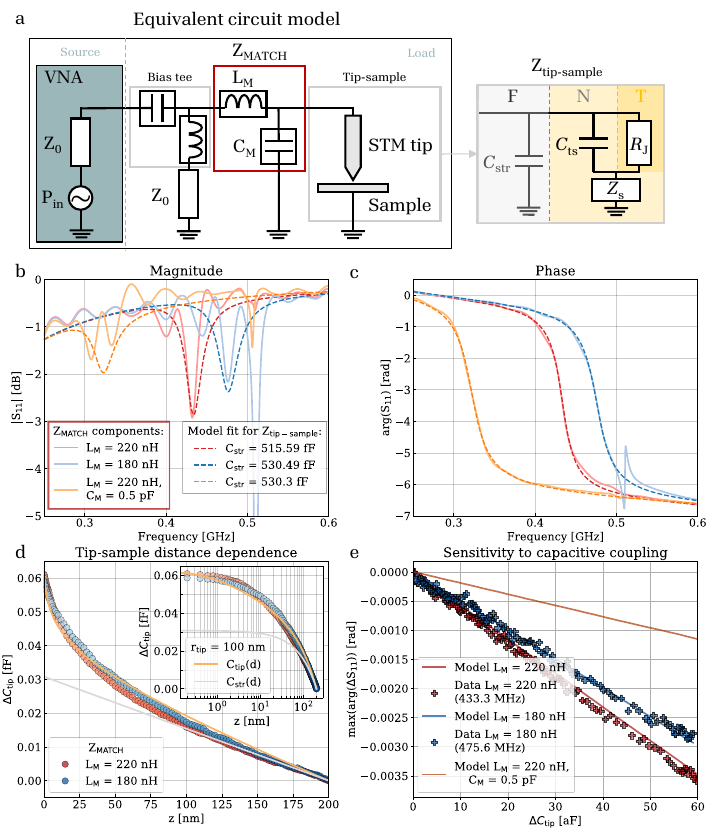}
    \caption{a) Equivalent circuit model seen from the source (VNA, in teal). The bias tee has capacitor C = 100 pF and inductor L = 120 nH. The matching circuit is different per chip. The equivalent circuit model of the tip-sample system depends on tip-sample distance. With stray capacitance $C_\mathrm{str}$ far away (F) from the sample, tip-sample capacitance $C_\mathrm{ts}$ in the near-field regime (N), and junction resistance $R_\mathrm{J}$ in tunneling (T). b,c) Measured S parameters ($P_\mathrm{in} = 0 \ \mathrm{dBm}$, $f_\mathrm{ifbw} = 1 \ \mathrm{kHz}$, average of 20 curves) and equivalent circuit model. d) Equivalent circuit model fit of the tip-sample distance dependence of the capacitive coupling, assuming $Z_s$ has a negligible contribution such that we have one fit parameter $C_\mathrm{tip}=C_\mathrm{str}+C_\mathrm{ts}$. We fit with a model for $C_\mathrm{str}$ and $C_\mathrm{ts}$, showing that the effective tip radius $R=100\ \mathrm{nm}$. e) Sensitivity of the phase of the S parameters to capacitive coupling in the near-field and stray regime from equivalent circuit model (solid lines) and from the measurement (markers). The frequency at which the change in the phase is maximum is given between brackets in the legend.}
    \label{fig:fig2}
\end{figure}

We replaced the connectors of the STM head with SSMCX connectors, which connect to $\sim 12 \ \mathrm{cm}$ of CW2040-3650F flexible copper wires from which the jacket is removed to lower the vibrational coupling to the STM head. These wires are thermalized at 4 K, where they are connected to $\sim2 \ \mathrm{m}$ silverplated cupronickel semirigids from Keycom (ULT-21) that run from the 4 K plate to room temperature and are thermally anchored at 77 K and 50 K. The signal from port 1 of the VNA is attenuated by 30 dB; the reflected signal is separated by a directional coupler, amplified by 40 dB, and fed to port 2. The measured transmission is therefore proportional to the reflection coefficient of the tip-chip, which we refer to as $S_\mathrm{11}$. The network analyzer is uncalibrated, because the helium level in the bath cryostat changes through the week and with it the cable dissipation. Because the wiring consists of 50 $\Omega$ lines, we remove the static cable background by dividing each measurement by a reference taken with a plain tip and no circuit; small impedance mismatches at the connectors leave the standing-wave ripple visible in \hyperref[fig:fig2]{Fig. \ref*{fig:fig2}b}. The stray resonance around 510 MHz is also visible without impedance matching circuit, and has been removed from the reference measurement by a cubic spline fit over a small frequency range (490-525 MHz) to prevent any interference with our model fit. This stray resonance is therefore unrelated to our matching circuits, most likely related to stray coupling of the STM cables  which are soldered to the tip contacts. No stray effects have been removed from the measurements with matching circuits, apart from handling the slow drift by subtracting the off-resonance signal, where the matching resonance barely contributes.

The equivalent circuit model includes all contributions of the tip chip and tip-sample system (\hyperref[fig:fig2]{Fig. \ref*{fig:fig2}a}). The VNA probes the load impedance $Z_\mathrm{L}$ through the reflection coefficient $S_\mathrm{11} = (Z_\mathrm{L}-Z_\mathrm{0})/(Z_\mathrm{L}+Z_\mathrm{0})$, with $Z_\mathrm{0} = 50 \ \Omega$ \cite{Pozar2012}. The bias tee consists of a 120 nH inductor and a 100 pF capacitor with a cutoff frequency near 50 MHz. The combined AC and DC signal pass through an impedance matching network consisting of a series inductor $L_\mathrm{M}$ (180 or 220 nH) and, in specific configurations, a parallel capacitor $C_\mathrm{M}$ (0.5 pF). The bias tee and matching inductors are ceramic high Q inductors from Coilcraft 0402HP series. In the model we include the DC resistance and skin effect of the inductor wire in series with the inductance, as well as a parallel parasitic capacitance calculated from the self resonance frequency as $C_\mathrm{p}=1/((2\pi f_0 )^2 L)$. We assume that the resistivity of the copper wire of the inductor lowers by a factor $\eta = 5.5$ under cryogenic conditions, similar to values found in Ref \cite{Patra2020}, and reduce the DC resistance and skin effect accordingly by a factor $\eta$ and $\sqrt{\eta}$ respectively. The bias tee and matching capacitors are from Agilent Technologies and included as ideal capacitors in the model. Furthermore, a small background slope in magnitude and phase can be explained by the conducting path from the bias tee capacitor to the matching inductor which couples to ground through the epoxy layer that glues the grounded spring to the chip, and is included in the model as a microstrip transmission line. The only varying parameter per tip-chip from this contribution is the thickness of the epoxy layer, whereas its width and length are fixed. In summary, our model includes 18 parameters of which only 2 parameters vary per chip; the stray capacitance and the epoxy layer thickness.

\section{Reactive channel}

We start far away from the sample where there is no junction, and the response is purely reactive. \hyperref[fig:fig2]{Fig. \ref*{fig:fig2}b} and \hyperref[fig:fig2]{c} show $S_\mathrm{11}$ for three matching circuits with the equivalent-circuit model. The impedances together give a dip in reflection at a matching frequency specific to each circuit. All three circuits are overcoupled: the phase winds by nearly 2$\pi$ through the resonance (\hyperref[fig:fig2]{Fig. \ref*{fig:fig2}c}), because the internal losses are smaller than the coupling to the line. At critical coupling, the reflection would vanish on resonance and the sensitivity would be largest. Comparing the circuits in \hyperref[fig:fig2]{Fig. \ref*{fig:fig2}b}, this configuration matches better for larger $L_\mathrm{M}$ and for smaller $C_\mathrm{M}$ and $C_\mathrm{str}$. Placing the matching circuit directly behind the tip pays off in the stray capacitance: from the off-sample data we fit $C_\mathrm{str}\approx 0.5 \ \mathrm{pF}$ for all three chips, two to twenty times smaller than the 1 to 10 pF usually quoted for microwave scanning probes \cite{Voogd2017,Barber2022,Kemiktarak2007,Bastiaans2018}. As the tip approaches, the capacitance to ground grows and pushes the matching resonance to lower frequency.

Assuming the sample impedance is negligible here, a single free parameter, the tip capacitance $C_\mathrm{tip}=C_\mathrm{str}+C_\mathrm{ts}$, describes the whole approach curve in the near-field and stray regimes (\hyperref[fig:fig2]{Fig. \ref*{fig:fig2}d}). The fits give the same tip capacitance versus height for two different matching circuits (red and blue). The model of Ref. \cite{Voogd2017}, with $C_\mathrm{str} = \varepsilon_0A/(L+d)$ and $C_\mathrm{ts} =  2\pi\varepsilon_0r_\mathrm{tip}\ln{(8r_\mathrm{tip}/d)}$, reproduces this result for a tip with length $L =  3.40 \ \mathrm{mm}$ long and $A =  201 \ \mathrm{mm}^2$ with an effective radius $r_\mathrm{tip} = 100 \ \mathrm{nm}$, so the reactive channel yields the effective tip radius. The circuits differ in how strongly they respond to this channel. The model predicts that the phase near the resonance frequency changes linearly with capacitance, with a slope of $-5.80 \times 10^{-5}$ rad/aF for the 220 nH circuit and $-4.97 \times 10^{-5}$ rad/aF for the 180 nH circuit, and it drops to $-1.92 \times 10^{-5}$ rad/aF once a parallel capacitor $C_\mathrm{M}$ is added, because the larger parallel capacitance dilutes the contribution of $C_\mathrm{ts}$. The measurements on the two circuits without $C_\mathrm{M}$ agree with this prediction (\hyperref[fig:fig2]{Fig. \ref*{fig:fig2}e}). Hence, the replaceable circuits set the sensitivity to the reactive channel in a controlled way. 

\section{Dissipative channel}

\begin{figure}
    \centering
    \includegraphics[width=\linewidth]{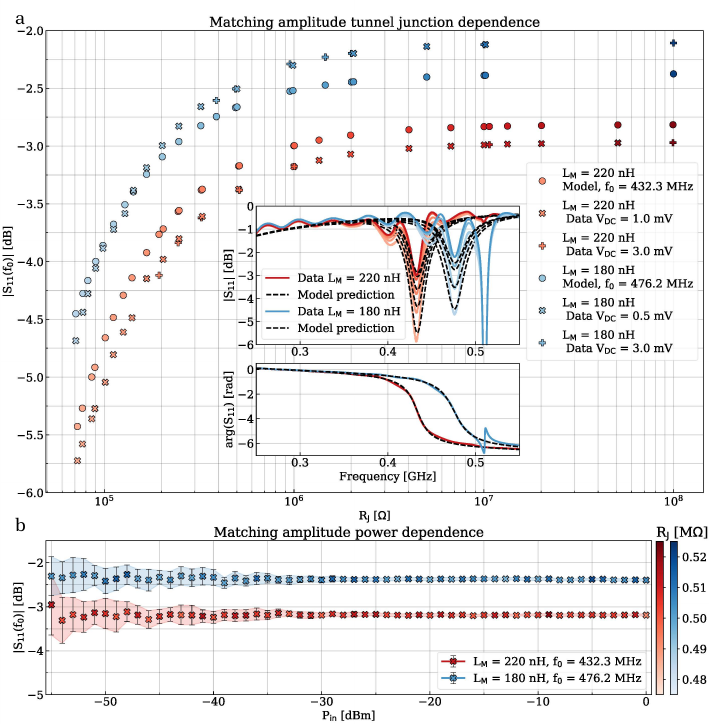}
    \caption{a) Tip-sample tunnel junction dependence model prediction and data ($V_\mathrm{DC}$ in legend). Inset shows the S parameters of the model (black broken curves) and data (colored full curves, $P_\mathrm{in} = 0 \ \mathrm{dBm}$, $f_\mathrm{ifbw} = 1 \ \mathrm{kHz}$, average of 20 curves) for several values of $R_\mathrm{J}$. b) Matching amplitude on resonance as a function of input power ($V_\mathrm{DC} = 10 \ \mathrm{mV}$, $f_\mathrm{ifbw} = 1 \ \mathrm{kHz}$, $R_\mathrm{J} = 0.5 \ \mathrm{M}\Omega$). The measurement is repeated 20 times and the errorbars are determined from the standard deviation.}
    \label{fig:fig3}
\end{figure}

Bringing the tip into tunneling adds the dissipative resistance $R_\mathrm{J}$ in parallel with $C_\mathrm{ts}$. Unlike the capacitance, $R_\mathrm{J}$ mainly changes the depth of the matching resonance rather than its frequency (inset in \hyperref[fig:fig3]{Fig. \ref*{fig:fig3}a}), so the two channels appear largely in different features of the same $S_\mathrm{11}$; a small cross-coupling remains. We treat the junction under three assumptions: that it responds linearly, so we ignore any modulation of $R_\mathrm{J}$ by the microwave voltage; that its microwave resistance equals the DC value $R_\mathrm{J}=V_\mathrm{DC}/I_\mathrm{tun}$, which holds for the nearly ohmic Au(111) junction at the small biases used here; and that the capacitances stay fixed over the tunneling range. The third assumption follows from our capacitive-regime measurements: $C_\mathrm{ts}$ grows by only 3.6 aF between $R_\mathrm{J}$ = 100 $\mathrm{M}\Omega$ and 0.1 $\mathrm{M}\Omega$, which moves the phase by at most $2.1 \times 10^{-4}$ rad and the magnitude by at most $1.6 \times 10^{-4}$ dB. This is more than three orders of magnitude smaller than the changes the junction resistance produces (\hyperref[fig:fig3]{Fig. \ref*{fig:fig3}a}). We define $f_0$ as the matching resonance frequency, where the dip depth is largest. This is also the frequency where the contrast with respect to changes in $R_\mathrm{J}$ is most visible. Each time the tip-chip is mounted in the STM head, this frequency shifts slightly, due variations in stray capacitance of $\sim$10s of aF upon reinserting the tip-chip. During the VNA measurements, the STM feedback is frozen so that the tip-sample distance does not change. The DC value of $R_\mathrm{J}$ is measured without any microwave input power $P_\mathrm{in}$.

With these assumptions, the model predicts the change in $S_\mathrm{11}$ caused by a change in $R_\mathrm{J}$ without additional free parameters: the circuit parameters and $C_\mathrm{str}$ are fixed by the off-sample data, and $R_\mathrm{J}$ is set by the DC setpoint. The model follows the measured matching amplitude against $R_\mathrm{J}$ for both circuits (\hyperref[fig:fig3]{Fig. \ref*{fig:fig3}a}), with better impedance matching at lower junction resistance: lowering $R_\mathrm{J}$ adds loss and brings the overcoupled circuits closer to critical coupling. $R_\mathrm{J}$ loads the matching circuit and at large $R_\mathrm{J}$ it barely perturbs it, leaving the external coupling dominant. As $R_\mathrm{J}$ decreases, it starts to short out the matching resonance and increases the internal loss channel. Above a few M$\Omega$ the matching amplitude saturates, so the present circuits are most sensitive at low junction resistance. On Au(111), the trend does not depend on which DC bias and current give a particular $R_\mathrm{J}$ (tested between 0.5 and 3 mV), nor on the microwave input power $P_\mathrm{in}$ between -55 and 0 dBm (\hyperref[fig:fig3]{Fig. \ref*{fig:fig3}b}). This means the DC setpoint sets the microwave junction impedance. Both circuits follow changes in the junction resistance. The matched circuit therefore reads out the dissipative tunneling channel at microwave frequencies, and for both circuits we can predict its sensitivity and interpret a change in $S_\mathrm{11}$ quantitatively as a change in tip–sample impedance, once the stray capacitance has been determined off sample.

\begin{figure}
    \centering
    \includegraphics[width=\linewidth]{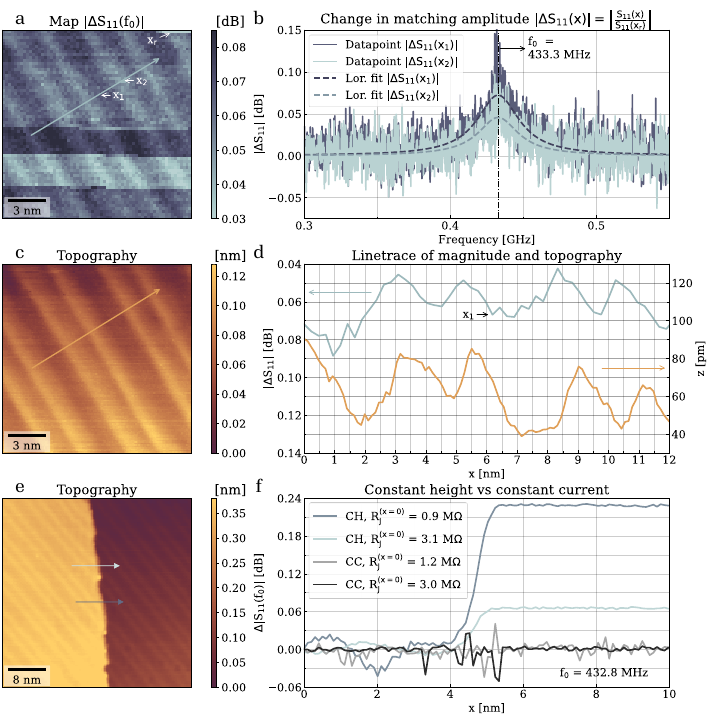}
    \caption{a) Spatial dependence of matching amplitude on resonance ($L_\mathrm{M}$ = 220 nH) in constant height with DC voltage line disconnected ($P_\mathrm{in} = -30 \ \mathrm{dBm}$ without attenuators, $f_\mathrm{ifbw} = 1 \ \mathrm{kHz}$, $\Delta x = 0.24 \ \mathrm{nm}$, $R_\mathrm{J} \approx 1 \ \mathrm{M}\Omega$), showing atomic resolution. The data has been normalized with respect to pixel $x_r$, and the change in magnitude is obtained by fitting the normalized data with a lorentzian. b) Data and lorentzian fit at pixels $\mathrm{x}_1$ and $\mathrm{x}_2$, normalized with respect to pixel $\mathrm{x}_\mathrm{r}$. c) STM topography ($V_\mathrm{DC} = -3\ \mathrm{mV}$, $I_\mathrm{tun} = -50 \ \mathrm{pA}$) in constant current mode, taken before the map in b at the same location. d) Linetrace from map (blue) and topography (orange) at the location indicated by the arrow in a and c respectively. e) Topography of step edge ($V_\mathrm{DC} = 50 \ \mathrm{mV}$, $I_\mathrm{tun} = 20 \ \mathrm{pA}$). Arrow indicates the location of the linetraces in f. f) Matching amplitude linetraces in constant height (CH) and constant current (CC) ($L_\mathrm{M}$ = 220 nH). Measurement settings: $V_\mathrm{DC} = 6.0 \ \mathrm{mV}$, $P_\mathrm{in} = 0 \ \mathrm{dBm}$, averaged 3 times, $f_\mathrm{ifbw} = 1 \ \mathrm{kHz}$, $\Delta x = 0.1 \ \mathrm{nm}$.}
    \label{fig:fig4}
\end{figure}

\section{Microwave-only imaging}
An important check is whether the dissipative channel resolves atomic-scale height changes. The capacitive channel averages over the tip radius, which is significantly larger then the atomic scale, whereas $R_\mathrm{J}$ is set by the last atom of the tip and depends exponentially on distance. \hyperref[fig:fig4]{Fig. \ref*{fig:fig4}a} is a constant-height map of the matching amplitude taken without DC bias, so no DC tunneling current flows and only the microwave voltage present in the junction. The map was setup at $R_\mathrm{J} = 1 \ \mathrm{M}\Omega$, after which the feedback is frozen and the bias line was physically disconnected. In this measurement, the cabling setup is different from \hyperref[fig:fig1]{Fig. \ref*{fig:fig1}a}, with Coax 1 directly connected to port 1, instead of separating the input and output lines, and without the 30 dB attenuators. The map shows the Au(111) herringbone reconstruction \cite{Barth1990}, with a corrugation of a few tens of picometers, and matches the DC topography recorded beforehand on the same region (\hyperref[fig:fig4]{Fig. \ref*{fig:fig4}c}, \hyperref[fig:fig4]{d}). The microwave channel thus resolves tip–sample distance changes of about 20 pm, corresponding to about 0.015 dB in matching amplitude. 

To confirm that the contrast comes from the junction resistance and not from the reactive channel, we take line scans across a monatomic step on the Au(111) surface, 0.24 nm high (\hyperref[fig:fig4]{Fig. \ref*{fig:fig4}e}), at two junction setpoints (\hyperref[fig:fig4]{Fig. \ref*{fig:fig4}f}). In constant height, the step raises $R_\mathrm{J}$ by about two orders of magnitude. Starting from $R_\mathrm{J} = 0.9 \ \mathrm{M}\Omega$, the matching amplitude changes by 0.229 dB; starting from $3.1 \ \mathrm{M}\Omega$, it changes by only 0.066 dB. The model predicts changes of 0.209 and 0.060 dB respectively for the two setpoints. This setpoint dependence follows the junction-resistance dependence of \hyperref[fig:fig3]{Fig. \ref*{fig:fig3}a}, whose slope, and therefore the sensitivity, is larger at smaller $R_\mathrm{J}$. The capacitive channel cannot account for it: over the entire tunneling range,  $C_\mathrm{ts}$ changes by only 3.6 aF according to our model, which moves the magnitude by at most $1.6 \times 10^{-4}$ dB, more than three orders of magnitude less than observed, and this change would not depend on the setpoint. In constant current, where the feedback holds $R_\mathrm{J}$ fixed and the tip follows the surface, the contrast disappears. Thus, the contrast does not come from the reactive, capacitive response that dominates near-field microwave microscopy but from the dissipative tunnel junction, read out through the matched circuit. Because the present circuits are overcoupled, a circuit closer to critical coupling would increase the contrast further.

\section{Conclusions}
In summary, we have built a microwave STM with in situ replaceable tip-chips that place the matching circuit directly behind the tip, and used it to read out both the reactive and the dissipative parts of the tip-sample load. A single equivalent-circuit model describes all circuits with two chip-specific parameters, yields the effective tip radius from the reactive channel, and predicts the response to the junction resistance. Read out through the matched circuit, the tunnel junction images Au(111) with about 20~pm height sensitivity from the microwave signal alone.

As a caveat, we have not yet extracted the sample impedance $Z_\mathrm{s}$ itself. For superconductors, we can make this impedance concrete. As shown in the Appendix, using quasiclassical theory for a normal tip tunneling into a dirty superconductor \cite{Kupriyanov1988,Belzig1999}, we find that for $\hbar\omega,\, k_\mathrm{B}T \ll \Delta$, with $\Delta$ the superconducting gap, single-electron tunneling is blocked and the junction turns from a resistance into a capacitance $C_\mathrm{dyn} = \hbar/(2R_\mathrm{J}\Delta)$, where $R_\mathrm{J}$ is the normal-state junction resistance. This is the same zero-bias, linear-response configuration as in \hyperref[fig:fig4]{Fig. \ref*{fig:fig4}a}. In series with the junction, the sample adds a resistance, because the injected quasiparticle current converts into supercurrent only within about a coherence length $\xi$ of the tip, and a kinetic inductance $L_\mathrm{k}$ set by the phase stiffness of the condensate. Notably, $L_\mathrm{k}$ depends on $\xi$ rather than on the tip radius; for a bulk sample, $L_\mathrm{k} \approx \hbar/(2\pi^2 \sigma_\mathrm{D} \Delta \xi)$, with $\sigma_\mathrm{D}$ the normal-state conductivity. Both terms are most sensitive to the superfluid weight close to the tip.

Read out through the tunnel junction, the superconducting response would thus be resolved on the scale of $\xi$, nanometers to tens of nanometers, rather than on the 100~nm scale of the capacitive channel. We emphasize that these sample terms are small compared with the junction impedance at the setpoints used here, and that $k_\mathrm{B}T \ll \Delta$ requires temperatures well below 4.2~K for most superconductors. At finite bias, the microwave drive also modulates the junction, whose response becomes a photon-assisted (Tien--Gordon) average \cite{Tien1963,Peters2020} that a quantitative analysis will have to include.

A natural next step is a measurement in which the STM feedback fixes $R_\mathrm{J}$, and with it the tip--sample distance to exponential precision, so that any remaining change in impedance can be assigned to the local sample response. Because the matching circuit can be replaced in situ, it can be tailored to the impedance range of each sample, whereas existing microwave impedance microscopes are usually limited to the range set by a fixed matching circuit. Local maps of the complex impedance, and of the kinetic inductance connected to quantum-geometry effects, could then help answer whether macroscopic superconductivity in inhomogeneous and unconventional superconductors is suppressed by weak pairing or by a loss of phase stiffness.

\acknowledgments

This project was supported by the Netherlands Organisation for Scientific Research (NWO/OCW), as part of the Frontiers of Nanoscience program (No. NF21CAS18). This work was supported by a collaboration between The Kavli Foundation, K. Tschira Stiftung, and K. Wells as part of the SuperC collaboration. This work was supported by the European Research Council (ERC StG SpinMelt, grant agreement 758572) and by the European Union (ERC, PairNoise, 101043833). The work of T. K. and T. T. H. was supported by the Research Council of Finland through DYNCOR, Project Number 354735. The work was done within the SuperC collaboration, supported by the Keele Foundation and the Jane and Aatos Erkko Foundation, and within the Finnish Centre of Excellence in Quantum Materials (QMAT). S. M. A. acknowledges funding by the U S Department of Energy/High Energy Physics through Grant No. DESC0017931, and the Army Research Office under Grant Number ARO W911NF-24-1-0153. We acknowledge funding by the Casimir PhD Travel Grant.

\bibliography{biblio.bib}
\appendix
\onecolumngrid\
\section{Derivation of RLC model for STM - superconductor junctions}

In this appendix we consider the impedance of the tip and sample. We show that the tip-sample contact is characterized by a dynamical capacitance, the quasiparticle to current conservation provides a resistance, while the supercurrent response introduces an inductance.

\begin{table}[ht]
    \centering
    \begin{tabular}{|c|c|c|}
    \hline
          & 2D&3D\\
         \hline
         C$_{\text{dyn}}$& $\frac{\hbar}{2R_\mathrm{J}\Delta}$ &$\frac{\hbar}{2R_\mathrm{J}\Delta}$ \\
         R&$\frac{1}{\pi\sigma_D}\text{ln}\frac{\xi}{r_t}$&$\frac{1}{2\pi^2\sigma_D r_t}$\\
         L& $\frac{\hbar}{\pi\sigma_D\Delta}\text{ln}\frac{r_s}{\xi}$&$\frac{1}{2\pi^2\sigma_D \Delta \xi}$\\
         \hline
    \end{tabular}
    \caption{The dynamical capacitance of the tip, the resistance and the inductance of the STM / superconductor junction.}
    \label{tab:parametersummary}
\end{table}

Due to the finite length required for quasiparticle to condensate conversion, the inductance does not depend on the radius of the tip  $r_t$, but on the superconductor coherence length $\xi$ and the size of the sample $r_s$. The scale of the tip radius enters in the dissipative component of the junction, which increases for smaller tip sizes because of the high current densities at small $r$.  The expressions for the different components of the impedance are summarized in Table \ref{tab:parametersummary}.

Subsequently, we consider the change of the response of the impedance due to variations in the conductivity. The dissipative channel is most sensitive to local perturbations of the superfluid weight, with perturbations at a distance $r$ being weighted as $\sim \frac{1}{r^2}$, in 2D and $\sim \frac{1}{r^4}$ in 3D.  

Moreover, even though the response in the reactive channel is inherently non-local because of the finite length required to convert the quasiparticle current into a supercurrent, it is most sensitive to the local superfluid weight. Its resolution however is worse than for the dissipative channels, with perturbations at a distance $r$ having a relative weight of $\sim \text{ln}(\frac{r}{\xi})$ in 2D, and  $\sim 1/r^2$
 in 3D, summarized in Table \ref{tab:resolution}.
\begin{table}[ht]
    \centering
    \begin{tabular}{|c|c|c|}
    \hline 
    & 2D&3D\\
      R& $r^{-2}$  & $r^{-4}$ \\
       L& $-\text{ln}(r)$ & $r^{-2}$\\
       \hline
    \end{tabular}
    \caption{Relative sensitivity of different components to changes in the diffusion constant a distance $r$ from the tip. }
    \label{tab:resolution}
\end{table}

Thus, by scanning over the surface, one obtains information about the superfluid weight distribution in two ways, both via variations in the resistive and in the reactive part of the impedance.

\subsection{Current}
We consider a tunneling contact with radius $r_t$ that is close to a superconductor. We consider the tip to be in the center of a superconductor that is either a 1D chain, a 2D slab or a 3D sample, and denote the distance between the tip and the sample edge $r_s$. 

We first focus on the impedance of the sample of the tip-sample interface. We assume there is no additional DC bias in addition to the AC signal. Thus, we consider a voltage drop $V_b e^{-i\omega t}$ and calculate the current between the STM tip and the part of the sample right below it. We describe the superconductor using the quasiclassical formalism, assuming that the superconducting gap is much smaller than the Fermi energy. Because STM systems are in the tunneling regime, we can use the Kuprianov-Luckichev boundary conditions \cite{Kupriyanov1988} to describe the boundary. Moreover, to lowest order we ignore proximity and inverse proximity effects. Therefore, the tip is in the normal state, $g_N(t,t') = e^{-iA(t)/2\tau_3}g_{N}e^{iA(t')/2\tau_3}$ where $A(t) = \frac{V_b}{\omega}$ and $g_{N}$ is the quasiclassical Green's function of the normal metal, that is, $g_{N}^R = -g_{N}^A = \tau_3$ and $g_{N}^K = 2\tau_3\tanh\frac{E}{2k_BT}$, where $\tau_3$ is the third Pauli matrix in Nambu space. For the superconductor we use the standard \textit{s}-wave quasiclassical Green's function. Denoting convolutions in time by $\circ$, we obtain 
\begin{align}
    I &= \frac{1}{8R_\mathrm{J}}\text{tr}\rho_1 \tau_3 [[A\tau_3\commast g_N]\commast g_S]|_{t = t'}\nonumber\\& = \int dE \frac{1}{4R_\mathrm{J}}V\text{tr}\rho_1 \tau_3 \Big(\tau_3g_N(E+\omega)g_S(E+\omega)-g_N(E)\tau_3g_S(E+\omega)-g_S(E)\tau_3g_N(E+\omega)+g_S(E)g_N(E)\tau_3\Big)\nonumber\\&
    = \frac{1}{8 R_\mathrm{J}\omega}V_b\int dE \text{tr}\Big((g^R_N(E+\omega)-g^R_N(E))g^K_S(E+\omega)-g^R_S(E)(g^K_N(E+\omega)-g^K_N(E))\nonumber\\&+(g^K_N(E+\omega)-g^K_N(E))g^A_S(E+\omega)-g^K_S(E)(g^A_N(E+\omega)-g^A_N(E))\Big)\;,\label{eq:KL}
\end{align}
Using that $g^R_N$ and $g^A_N$ are independent of energy, we find
\begin{align}
    I&= \frac{1}{8 R_\mathrm{J}\omega}V_b\int dE \text{tr}\tau_3(g^A_S(E+\omega)-g^R_S(E))(f(E+\omega)-f(E))
\end{align}
where $f(E) = \tanh{\frac{E}{2k_B T}}$ is the equilibrium distribution function. At zero temperature $f(E) = \text{sign}(E)$, so that this reduces to 
\begin{align}
    I & = \frac{1}{8 R_\mathrm{J}\omega}V_b \int_{-\omega}^{0} dE \text{tr}\Big(\tau_3 (g^A_S(E+\omega)-g^R_S(E))\Big)\;.
\end{align}
For frequencies below the gap, this reduces to 
\begin{align}\label{eq:CurrentExpression}
    I = \frac{V_b}{R_\mathrm{J}} \frac{1}{\omega}\int_{-\omega}^{0} dE \frac{-iE}{\sqrt{\Delta^2-E^2}}dE = \frac{V_b}{R_\mathrm{J}}\frac{i}{\omega}(\Delta-\sqrt{\Delta^2-\omega^2})\;. 
\end{align}
As expected, for the normal state ($\Delta\xrightarrow{}0$) this converges to $\frac{V_b}{R_\mathrm{J}}$. For the superconducting state on the other hand there is no DC current to first order in tunneling. Thus, for small frequencies ($\omega\ll \Delta$) $I\approx \omega \frac{V_b}{2R_\mathrm{J} \Delta}$, corresponding to capacitive behavior of the interface. 

\subsection{Voltage drops inside the superconductor}
We have now related the voltage drop between the tip and the sample to the current through the tip. However, there is an additional contribution to the impedance of the sample itself. In this section we calculate the voltage drop inside the superconductor, and show that the voltage can be separated in a component proportional to the current, i.e. a resistance, and a voltage proportional to the time-derivative, i.e. an inductance. 

We perform perturbation theory in the magnitude of the current. Therefore, we choose our gauge such that it contains all of the superfluid momentum, that is, the superconducting phase is constant. We assume that the superconductor is dirty, $\frac{\hbar}{\tau}\gg \Delta$, so that we can employ the Usadel equation. 

For the charge current we have, for small frequencies  \cite{Belzig1999}
\begin{align}
    \nabla\cdot j &= \frac{1}{8}\text{tr}[g,\hat{\Delta}]\;,\label{eq:continuity}\\
    j(R = r_0) &= \frac{\mathcal{I}(E)}{A_d}\;,\\
    j &= D_T\nabla f_T+\text{ImIs} f_L\;,
\end{align}
Here $D_{T} = \frac{1}{4}\text{tr}(1-g^R\tau_3g^A\tau_3)$ is the diffusion constant for charge, $A_d$ is the dimension ($d$) dependent area ($A_1 = 2$, $A_2 = \pi r_t$, $A_3 = 2\pi r_t^2$). The spectral current $\mathcal{I}(E) = \frac{V_b}{2R_\mathrm{J}}\frac{1}{i\omega}(\frac{E}{\sqrt{\Delta^2-E^2}} - \frac{E+\omega}{\sqrt{\Delta^2-(E+\omega)^2}})$ is the integrand of the current in Eq. (\ref{eq:CurrentExpression}), while $\text{ImIs} =  \text{tr}\tau_3(g^R\nabla g^R-g^A\nabla g^A)$.

To solve these equations, we exploit features of the retarded and advanced equation. As noted in Ref. \cite{Belzig1999}, the continuity equation, Eq. (\ref{eq:continuity}) can be simplified using that the retarded and advanced Green's function satisfy the Usadel equation, so that Eq. (\ref{eq:continuity}) reduces to
\begin{align}
    \nabla (D_T \nabla f_T) + \text{ImIs}\nabla f_L = \mathcal{R}f_T
\end{align}
where the relaxation rate $\mathcal{R} = \frac{1}{8}\text{tr}\hat{\Delta}(g^R+g^A) + \frac{i\omega}{8}\text{tr}\tau_3(g^R-g^A)$ contains terms stemming from the quasiparticle-condensate conversion (via $\hat{\Delta}$) and from the time-derivative term.

If we consider the problem up to linear order in voltage, the problem simplifies. Because in the absence of a voltage both $\text{ImIs}$ and $\nabla f_L$ vanish, we know that this term is at least second order in voltage and therefore it drops out. Moreover, for $D_T$ and $\mathcal{R}$ we may take their unperturbed BCS values,
\begin{align}D_{T} &= D\Big(1+\frac{\Delta^2+E(E+\omega)}{\sqrt{\Delta^2-E^2}\sqrt{\Delta^2-(E+\omega)^2}}\Big)\;,\\\mathcal{R} &= \frac{1}{2}\Big(\frac{\Delta^2}{\sqrt{\Delta^2-E^2}}+\frac{\Delta^2}{\sqrt{\Delta^2-(E+\omega)^2}} = \frac{\Delta^2(\sqrt{\Delta^2-E^2}+\sqrt{\Delta^2-(E+\omega)^2})}{\sqrt{\Delta^2-E^2}\sqrt{\Delta^2-(E+\omega)^2}}\Big).\end{align} With this, the bulk equation simplifies to
\begin{align}\label{eq:Keldyshbulk}
    \nabla^2 f_T = \frac{1}{\xi_E^2}f_T\;,
\end{align}
where $\xi_E^2 = \frac{D_{T0}}{\mathcal{R}_0} = \xi \frac{\Delta^2+E(E+\omega) + \sqrt{\Delta^2-E^2}\sqrt{\Delta^2-(E+\omega)^2}}{\Delta (\sqrt{\Delta^2-E^2}+\sqrt{\Delta^2-(E+\omega)^2})}$ is the effective energy dependent decay length for the transverse distribution function, and $\xi = \sqrt{\frac{D}{\Delta}}$ is the standard superconductor coherence length.

Since Eq.~(\ref{eq:Keldyshbulk}) is a diffusion equation with constant coefficients, it can be solved analytically. We have to supplement this equation with the boundary condition, Eq. (\ref{eq:CurrentExpression}).

Also the boundary condition can be simplified by exploiting that the retarded and advanced component satisfy the Usadel equation. Specifically, for energies below the gap  $g_S^R = g_S^A$. Therefore, for frequencies smaller than the superconductor gap, we find that at the boundary the retarded matrix current satisfies
\begin{align}
    I_R &= \text{tr}\tau_3\frac{1+\rho_3}{2}[[A\tau_3\commast g_N]\commast g_S]|_{t = t'} = \nonumber\\& = \int_{-\omega}^0 dE \frac{1}{4R_\mathrm{J}}V\text{tr}\rho_1 \tau_3 \Big(\tau_3g_N(E+\omega)g_S(E+\omega)-g_N(E)\tau_3g_S(E+\omega)-g_S(E)\tau_3g_N(E+\omega)+g_S(E)g_N(E)\tau_3\Big)\nonumber\\& = \int_{-\omega}^0 dE \frac{1}{4R_\mathrm{J}}V\text{tr}(g^R_N(E+\omega)g^R_S(E+\omega)-g^R_N(E)g^R_S(E+\omega)-g^R_S(E)g^R_N(E+\omega)+g^R_S(E)g^R_N(E)) = \int dE 0 = 0\;,
\end{align}
where we used again that $g^R_N$ is energy independent. This means that $\text{ImIs}$ vanishes at the boundary, that is, at the boundary there is no supercurrent all current enters through the non-equilibrium distribution function and $j(r = r_t) = D_T\nabla f_T$. This reduces the boundary condition, for the charge current, Eq.~(\ref{eq:KeldyshBC}), to
\begin{align}\label{eq:KeldyshBC}
    D_T \nabla f_T = -\frac{1}{  A_d}\frac{1}{\omega}\mathcal{I}(E)\;.
\end{align}

Using that $f_T$ must remain finite as $r\xrightarrow{}\infty$, the solution to this equation reads.
\begin{align}
    f_T(E) &= -\frac{\mathcal{I}(E) \xi_E}{D_T   A_d} e^{-(r-r_t)/\xi_E}\;, & 1D\;,\label{eq:fT1D}\\
    f_T(E) & = \frac{\mathcal{I}(E) \xi_E}{D_T   A_d} \frac{K_0(\frac{r}{\xi_E})}{K_1 (\frac{r_t}{\xi_E})}\;, & 2D\;,\\
    f_T(E) &= \frac{\mathcal{I}(E) \xi_E}{D_T   A_d} \frac{r_t}{r_t+\xi_E}\frac{r_t}{r}e^{-(r-r_t)/\xi_E}\;,&3D\;.
\end{align}

With this, the quasiparticle current  density $j_q$ can be written as
\begin{align}
    j_q &= \frac{1}{\omega}\int_{-\omega}^{0} dE \frac{\mathcal{I}(E)}{  A_d}e^{-|x|/\xi_E}\;, &1D\;,\\
    j_q & = \frac{1}{\omega}\int_{-\omega}^{0} dE \frac{\mathcal{I}(E)}{  A_d}\frac{K_1(\frac{r}{\xi_E})}{K_1(\frac{r_t}{\xi_E})}\;,&2D\;,\\
    j_q & = \frac{1}{\omega}\int_{-\omega}^{0} dE \frac{\mathcal{I}(E)}{   A_d} \frac{r_t^2}{r_t+\xi_E}\frac{r+\xi_E}{r^2}e^{-(r-r_t)/\xi_E}\;,&3D\;.
\end{align}
For small frequencies, we may estimate $\xi_E \approx\xi$ and $D_T \approx D$, so that the energy integral can be evaluated analytically. The result is
\begin{align}
    j_q &= \frac{I}{  A_d} e^{-\frac{r}{\xi}}\;,&1D\;,\\
    j_q &= \frac{I}{  A_d} \frac{K_1(\frac{r}{\xi})}{K_1(\frac{r_t}{\xi})}\;,&2D\;,\\
    j_q &= \frac{I}{  A_d}\frac{r_t^2}{r_t+\xi}\frac{r+\xi}{r^2} e^{-\frac{r}{\xi}}\;,&3D\label{eq:jq3D}\;,\\
\end{align}
where $I = \frac{V_b}{R_\mathrm{J}}\frac{i}{\omega}(\Delta-\sqrt{\Delta^2-\omega^2})$.

There are two important consequences to be drawn from Eqs.~(\ref{eq:fT1D}-\ref{eq:jq3D}). First of all, the transverse distribution function $f_T$ is position dependent. This means that the quasiparticles have have a different potential than the condensate close to the interface, which leads to a voltage drop.

This voltage drop is given by
\begin{align}
    V_R& = \int dE f_T(E,r = r_t)-\int dE f_T(E,r = r_s)\\
    &= \frac{I\xi}{2 \sigma_D }(1-e^{-\frac{r_s-r_t}{\xi}})\approx \frac{I\xi}{2\sigma_D}\;, & 1D\;.\\
    &= \frac{I\xi}{\pi \sigma_D r_t}\frac{K_0(\frac{r_t}{\xi})}{K_1(\frac{r_t}{\xi})} \approx \frac{I}{\pi \sigma_D}\text{ln}\frac{\xi}{r_t}\;, & 2D\;,\\
    &= \frac{I\xi}{2\pi\sigma_D r_t^2} \frac{r_t}{r_t + \xi} (1-e^{-\frac{r_s-r_t}{\xi}}) \approx \frac{I \pi \sigma_D}{r_t}\;, & 3D\;.
\end{align}
Thus, this voltage drop is proportional to the current, indicating a resistive response. 

Secondly, the quasiparticle current decays exponentially. However, the total charge current must be conserved, that is, the total current density $j$ decays as $\frac{1}{r^d}$.  Thus, the quasiparticle current is converted into a supercurrent, which reads $j_s(r) = j(r)-j_q(r) =j_q(r = r_t)\frac{r_t^d}{r^d}-j_q(r)$. 

Because it is a time-dependent supercurrent, this leads to a reactive voltage. To lowest order we have $j_s = \sigma_D \Delta q$, where $q$ is the superfluid momentum and $\sigma_D$ the conductivity. The voltage drop therefore reads
\begin{align}
    V_L &=  \int_{r_t}^{r_s} dr \partial_t \vec{A}\cdot\vec{\hat{r}} =i\frac{\hbar\omega}{e} \int_{r_t}^{r_{s}} q dr \\ &=\frac{i\hbar\omega I}{2 \sigma_D \Delta}\Big((r_s-r_t)+\xi (e^{\frac{-r_s}{\xi}}-e^{\frac{-r_t}{\xi}})\Big)\;,&1D\;,\\
    &= \frac{i\hbar\omega I}{\pi   \sigma_D \Delta}\Big(\text{ln}\frac{r_s}{r_t}+\frac{\xi}{r_t}\frac{K_0(\frac{r_s}{\xi})-K_{0}(\frac{r_t}{\xi})}{K_{1}(\frac{r_t}{\xi})}\Big)\;,&2D\;,\\
    &=\frac{i\hbar\omega I}{2\pi^2\sigma_D \Delta}\Big(\frac{1}{r_t}(1-e^{-\frac{r_t}{\xi}})-\frac{1}{r_s}(1-e^{-\frac{r_s}{\xi}})\Big)\;,&3D\;.
\end{align}
The ratio of the voltage drop $V_L$ and the current is linear in frequency. This corresponds to inductive behavior. For small tip radii, $r_t\ll \xi$, this reduces to
\begin{align}
    V_L &\approx \frac{i\hbar\omega I}{2 \sigma_D \Delta} (r_s-\xi)\;,& 1D\;,\\
    V_L &\approx \frac{i\hbar\omega I}{\pi \sigma_D \Delta} \text{ln}\frac{r_s}{\xi}\;,&2D\;,\\
    V_L &\approx \frac{i\hbar\omega I}{2\pi^2\sigma_D\Delta \xi}\;,&3D\;.
\end{align}
We conclude that the inductance does not depend on the tip radius $r_t$ but on the superconductor coherence length $\xi$. Summarizing the system, has a capacitance at the boundary, a resistance due to quasiparticle to supercurrent conversion within distance $\xi$ from the boundary, and an inductance in the rest of the sample.

\subsection{RLC parameters}
\begin{figure}
    \centering
    \includegraphics[width=\linewidth]{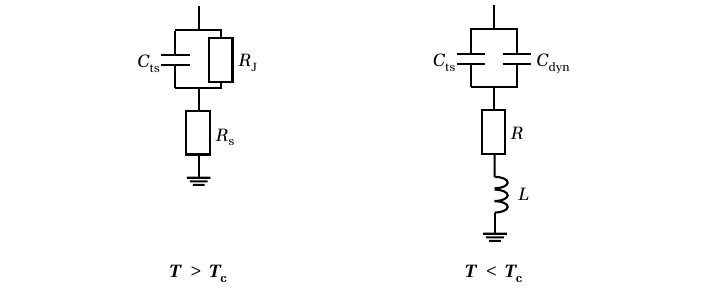}
    \caption{Sketch of the electronic circuit in the normal (left) and superconducting (right) states. }
    \label{fig:fig5}
\end{figure}

With this we have illuminated the relation between current and voltage in the different parts of the setup, which is summarized in the circuit diagram in \hyperref[fig:fig5]{Fig. \ref*{fig:fig5}}. Because the capacitance, resistance and inductance appear in series, it can be described using an RLC model. Here we summarize the RLC parameters of the system.

The capacitance has the same magnitude in all dimensions.
\begin{align}
    C_{\text{dyn}} = \frac{\hbar}{2R_\mathrm{J}\Delta}\;.
\end{align}
The resistance and inductance however depend strongly on the dimensionality.

In 1D
\begin{align}
    L&= \frac{V_L}{i\omega I} = \frac{\hbar}{2\sigma_D\Delta} (r_s-r_t+\xi(e^{-\frac{r_s}{\xi}}-e^{-\frac{r_t}{\xi}}) \approx \frac{\hbar}{2\sigma_D\Delta}(r_s-\xi)\;,\\
    R & = \frac{V_R}{I} = \frac{\xi}{2\sigma_D}\;,
\end{align}
and in 2D
\begin{align}
    L&= \frac{V_L}{i\omega I} =  \frac{\hbar}{\pi\sigma_D \Delta}\Big(\text{ln}\frac{r_s}{r_t}+\frac{\xi}{r_t}\frac{K_0(\frac{r_s}{\xi})-K_{0}(\frac{r_t}{\xi})}{K_{1}(\frac{r_t}{\xi})}\Big) \approx \frac{\hbar}{\pi\sigma_D \Delta}\text{ln}\frac{r_s}{\xi}\;,\\
    R&= \frac{V_R}{I} = \frac{\xi}{\pi \sigma_D r_t} \frac{K_{0}\left(\frac{r_t}{\xi}\right)}{K_1\left(\frac{r_t}{\xi}\right)}\approx \frac{1}{\pi \sigma_D}\text{ln}\frac{\xi}{r_t}\;,
\end{align}
and in 3D
\begin{align}
    L&= \frac{V_L}{i\omega I} = \frac{\hbar}{2\pi^2\sigma_D \Delta}\Big(\frac{1}{r_t}(1-e^{-\frac{r_t}{\xi}})-\frac{1}{r_s}(1-e^{-\frac{r_s}{\xi}})\Big)\approx \frac{\hbar}{2\pi^2\sigma_D\Delta \xi}\;,\\
    R&= \frac{V_R}{I} = \frac{r_t}{2\pi^2\sigma_D r_t^2 (r_t+\xi)} \approx \frac{1}{2\pi^2\sigma_D r_t}\;.
\end{align}

\subsection{Sensitivity to local perturbations in the diffusion constant}
Now that we have found the impedance for a homogeneous sample, we calculate how it changes upon small variations in the diffusion constant. In the preceding derivation we have found that the dissipative component is due to the quasiparticle flow close to the tip, while the inductive component is due to the supercurrents, which are generated over a distance $\xi$ from the tip. We show that because of this, the dissipative component has a better resolution than the reactive component.

First, we consider the effect on the quasiparticle to condensate conversion, and the change in the dissipative component that comes along with that. We again use the small frequency approximation $\omega\ll \Delta$, so that energy integrations can be performed analytically. To find the sensitivity of the dissipative component against local perturbations in the diffusion constant, we must solve the Usadel equation integrated over energy
\begin{align}
    \nabla (D_{T} + D_1\delta(\vec{r}_1)\nabla V)  = \frac{1}{\xi^2}V\;,
\end{align}
perturbatively in $D_1$. Denoting the perturbation in $V$ by $\delta V (\vec{\rho})$ where $\rho = \vec{r}-\vec{r}_1$, we find the equation
\begin{align}
    D_{T}\nabla^2 \delta V + D_1\nabla (\nabla V_0(r_1)\delta(\vec{r}_1)) = \frac{1}{\xi^2}\delta V\;.
\end{align}
In 2D this has the solution
\begin{align}
    \delta V(\rho) = \frac{D_1}{2\pi\xi}K_1 (\frac{\rho}{\xi})\frac{\vec{\rho}\cdot\nabla V_0(r_1)}{\rho}\;,
\end{align}
while in 3D
\begin{align}
    \delta V(\rho) = \frac{D_1}{2\pi^2}e^{-\frac{\rho}{\xi}}\frac{\rho + \xi}{\rho^2}\frac{\vec{\rho}\cdot\nabla V_0(r_1)}{\rho}\;.
\end{align}

Evaluating this at the tip ($\vec{r} = 0$ and $\vec{\rho} = -\vec{r}_1$), using that in 2D $\nabla V(\vec{r}_1) = -\frac{I}{2\pi r_t \sigma_D}\frac{\vec{r}_1}{r_1}$ while in 3D when $\vec{r}_1$ is on the surface $\nabla V(\vec{r}_1) = -\frac{I}{2\pi^2 r^2 \sigma_D}\frac{r_t^2}{r_t + \xi}\frac{r+\xi}{r^2}$\;, we find that
in 2D
\begin{align}
    \delta R = \delta V (\vec{r} = 0)/I = \frac{D_1}{4\pi^2\xi r_t K_{1}(\frac{r_t}{\xi})} (K_1(\frac{r_1}{\xi}))^2\;,
\end{align}
while in 3D
\begin{align}
    \delta R = \delta V (\vec{r} = 0)/I = \frac{D_1}{4\pi^4 \sigma_D}\frac{r_t^2}{r_t+\xi}\Big(\frac{r+\xi}{r^2}e^{-\frac{r}{\xi}}\Big)^2\;.
\end{align}
This gives the change in the dissipative component.
To solve for the change in the reactive component, we may use that $\Delta$ is constant to lowest order. Moreover, in the absence of a magnetic field, $\nabla\times\vec{q} = 0$, the superfluid momentum can be written as the gradient of a phase, $\vec{q} =\nabla\phi$. Therefore,  the supercurrent can be written as $\vec{j} = \sigma_D \nabla \phi$.

We define $\chi = V +  \phi$, and separate it as its value in a homogeneous system and a perturbation as $\chi = \chi_0 + \delta\chi$.
We find current conservation implies that its variation satisfies the equation
\begin{align}
    D_{T}\nabla^2 \delta \chi + D_1\nabla (\nabla \chi_0(r_1))\delta(\vec{r}_1) = 0\;.
\end{align}

This equation can be solved analytically. In 2D we find
\begin{align}
    \delta \chi(\vec{r} = 0) =  \frac{D_1}{4\pi^2}\frac{1}{r^2}\;,
\end{align}
while in 3D
\begin{align}
    \delta \chi(\vec{r} = 0) = \frac{D_1}{4\pi^4 \sigma_D}\frac{1}{r^4}\;.
\end{align}
Exploiting this, and using that all functions decay at large $r$, we find that in 2D
\begin{align}
    \delta L = \delta\phi/I = (\delta\chi-\delta V)/I =\frac{D_1}{4\pi^2} \Big(\frac{1}{r_1^2}-\frac{\xi}{r_t}K_{1}(\frac{r_t}{\xi})(K_{1}(\frac{r_1}{\xi})^2\Big)\approx\frac{D_1}{4\pi^2} \Big(\frac{1}{r_1^2}-(K_{1}(\frac{r_1}{\xi})^2\Big)\;,
\end{align}
while in 3D
\begin{align}
    \delta L = \frac{D_1}{4\pi^4 \sigma_D}\Bigg(\frac{1}{r^4}-\Big(\frac{r+\xi}{r^2}e^{-\frac{r}{\xi}}\Big)^2\Bigg)\;.
\end{align}
This shows that even though the inductance arises from phase gradients that are maximal a distance $\xi$ away from the tip, the reactances is most sensitive to the local perturbations of the superfluid weight.
\end{document}